\documentclass[reprint,aps,prb,superscriptaddress,nofootinbib,longbibliography]{revtex4-1}
\usepackage{braket}
\usepackage{times} 
\usepackage{xcolor}

\usepackage{amsfonts,amssymb,amsmath}
\usepackage[T1]{fontenc}

\usepackage{qcircuit}
\usepackage{xr-hyper}
\usepackage{hyperref}
\usepackage{enumitem}
\usepackage{comment}
\usepackage{makecell}
\usepackage[capitalize]{cleveref}
\usepackage{booktabs}
\usepackage{hhline}

\makeatletter
\newcommand*{\addFileDependency}[1]{
  \typeout{(#1)}
  \@addtofilelist{#1}
  \IfFileExists{#1}{}{\typeout{No file #1.}}
}
\makeatother

\usepackage{amsthm}

\newcommand*{\myexternaldocument}[1]{%
    \externaldocument{#1}%
    \addFileDependency{#1.tex}%
    \addFileDependency{#1.aux}%
}
\myexternaldocument{sup}
\hypersetup{colorlinks=true,citecolor=blue,linkcolor=blue, urlcolor=blue}
\hypersetup{linktocpage}
\newcolumntype{M}[1]{>{\centering\arraybackslash}m{#1}}
\newcolumntype{N}{@{}m{0pt}@{}}
\usepackage{environ}

\bibpunct{[}{]}{,}{n}{}{}

\usepackage{tikz}
\usetikzlibrary{arrows,snakes,backgrounds}

\myexternaldocument{sup}

\NewEnviron{eqs}{%
\begin{equation}\begin{split}
    \BODY
\end{split}\end{equation}
}

\begin{document}

\title{Quantum-Secured Device-Independent Global Positioning System}

\author{Chon-Fai Kam}
\affiliation{Department of Physics, University at Buffalo, SUNY, Buffalo, New York 14260, USA}

\author{En-Jui Kuo}
\affiliation{
Department of Electrophysics, National Yang Ming Chiao Tung University, Hsinchu, Taiwan, R.O.C.}

\begin{abstract}
This paper introduces a novel device-independent quantum self-testing protocol designed specifically for multipartite quantum communication. By exploiting the quantum rigidity in Bell nonlocality, the protocol enables the certification of genuinely entangled subspaces without reliance on device assumptions. Additionally, we investigate its potential to enhance the security of the Global Positioning System (GPS) against malicious cyberattacks. The study concludes with a comprehensive analysis of the experimental requirements, comparing superconducting and trapped-ion qubit architectures in terms of full-circuit fidelity and total gate time for generating a five-qubit code in the context of the noisy intermediate-scale quantum (NISQ) era.
\end{abstract}

\maketitle

\section{Introduction} 
Quantum communication marks a significant advancement in the future of technology, primarily due to its unparalleled security and rapid data transmission capabilities. Quantum Key Distribution (QKD) \cite{bennett2014quantum, ekert1991quantum} exemplifies this progress by offering encryption that is theoretically immune to cyber-attacks. Its foundation in quantum mechanics ensures that any attempt at eavesdropping disrupts the quantum state of particles, rendering the transmission invalid. This characteristic distinguishes it significantly from classical communication protocols. 

A notable class of quantum protocols that provides exceptionally robust protection for encrypted data is known as quantum device-independent protocols \cite{baccari2020device}. These protocols treat the encrypted quantum states as a ``black box" and verify their integrity in laboratories that are spatially separated through the application of Bell-type inequalities. According to the literature, both quantum secure direct communication (QSDC) \cite{long2002theoretically, deng2004secure} and quantum key distribution (QKD) \cite{acin2007device} have been demonstrated to incorporate device-independent protocols. As a result, these protocols are considered among the most advanced and secure technologies at the forefront of information security.

In this work, we examine a novel device-independent quantum communication protocol and demonstrate its applicability to protect the global positioning system (GPS) \cite{kaplan2017understanding} against cyberattacks. The protocol is centered on quantum self-testing, where a maximally entangled quantum state is shared between the satellite and the receiver. The rigidity of Bell non-locality shared between the satellites and receivers ensures that any tampering with the quantum state can be identified through the non-maximal violation of the Bell-CHSH-like inequalities. 

Recent studies \cite{sarkar2025using} have introduced protocols aimed at protecting the Global Positioning System (GPS) from simple attacks on transmissions between a satellite and a receiver, utilizing the bipartite Bell-CHSH inequality \cite{clauser1969proposed, bell1964einstein} to certify the maximally entangled Bell state shared between the two parties. However, a single satellite is insufficient for accurately determining the position of a three-dimensional object on Earth, as this requires at least three satellites. Moreover, a fourth satellite is essential to achieve high precision in measuring local time \cite{kaplan2017understanding}. Increasing the number of satellites further improves the accuracy of both time synchronization and distance measurement. As such, a multipartite quantum communication protocol is crucial for ensuring high-precision transmissions, such as those required by GPS, as it enables the secure and accurate exchange of information among multiple satellites and receivers. Intriguingly, Rigorous mathematical studies have also demonstrated that, in most cases, a network of five satellites is essential to pinpoint the exact location of a moving object on Earth with high precision \cite{boutin2024global}.

Central to this quantum protocol is the quantum self-testing of the five-qubit code, denoted as $[[5,1,3]]$, which was proposed by Baccari et al. \cite{baccari2020device}. This code encodes a single logical qubit into five physical qubits and represents the smallest possible code capable of correcting arbitrary single-qubit error. Intriguingly, this code facilitates a quantum communication protocol tailored for applications involving four satellites and a single receiver --- be it a car, an aircraft, or a ship --- where precise receiver positioning is critical. 

In this proposal, the quantum communication tasks between the satellites and the receiver are based on quantum self-testing of the entangled subspace within the physical code space. The structure of the entangled subspace ensures that errors affecting local physical qubits can be detected and corrected. Consequently, any hostile cyber attackers would need to compromise multiple physical qubits in a manner that exceeds the error correction capabilities of the code; otherwise, the attack would fail. Moreover, a direct extension of the Bell-CHSH inequality, like the Mermin inequality \cite{adhikari2016analytical, cabello2008mermin}, is inadequate for defending against all potential cyber attacks that introduce single-qubit errors on the physical qubits. 

Additionally, a simple quantum circuit for the five-qubit code, designed exclusively with two-qubit CNOT gates and single-qubit Hadamard gates, has been explicitly formulated \cite{mondal2024optimized}. This makes our quantum protocol particularly suited for quantum computers in the noisy intermediate-scale quantum (NISQ) era \cite{preskill2018quantum, brooks2019beyond}, where the fidelities of two-qubit gates and single-qubit states have improved to nearly or surpassing 99.99\%, with operation times on the scale of tens of nanoseconds for superconducting quantum systems. Also, to the best of our knowledge, the multipartite generalization of entangled states used in Bell-CHSH-type inequalities --- such as the Mermin-Ardehali-Belinskii-Klyshko (MABK) inequality \cite{panwar2023elegant}, Compact Bell Inequalities \cite{wu2013compact}, or Iterative Multipartite CHSH-Type Inequalities \cite{fan2023generalized} --- as well as most stabilizer codes, excluding the simplest Mermin inequality and the five-qubit code utilized in this work, lacks explicit quantum circuit designs. This underscores the unique significance of our multipartite quantum communication proposal.

Notice that our multipartite quantum communication protocol is not confined to the five-qubit code; more advanced error correction codes can be implemented for higher-precision scenarios involving more than five satellites, as long as the rigidity of quantum locality is maintained. Furthermore, our protocol differs from quantum key distribution methods, such as the BB84 protocol \cite{bennett2014quantum}, in that it does not require distinguishing between public and private keys. Additionally, we eliminate the need for classical channels, ensuring that the entangled states shared among the satellites remain secure and immune to compromise or jamming.

\begin{figure}
    \centering
\includegraphics[width=0.6\linewidth]{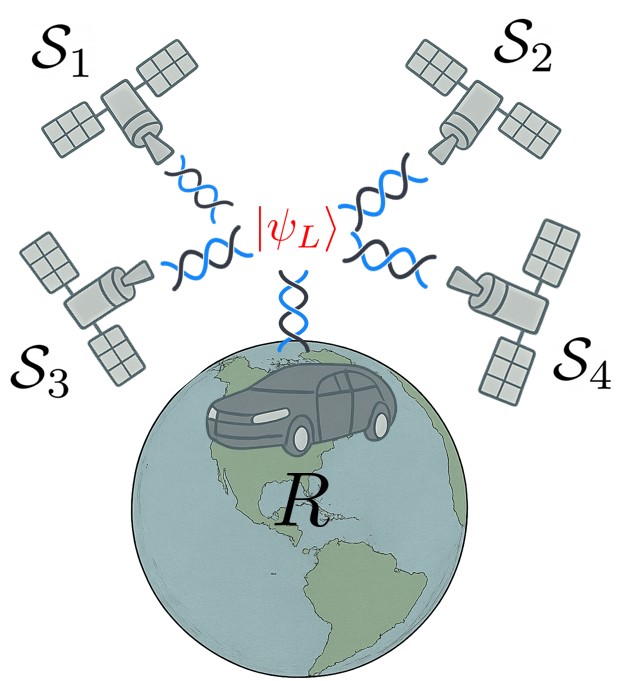}
    \caption{Illustration of a quantum-secured device-independent global positioning system. Here, we denote the four satellites as $\mathcal{S}_i$ for $i = 1, 2, 3, 4$, and the receiver as $R$. The middle state $\ket{\psi}$ illustrates the entangled states shared by $5$ parties.}
    \label{fig:enter-label}
\end{figure}

\section{Methods} 
At the core of our protocol is the principle of quantum self-testing. Quantum self-testing is a method that employs spacelike-separated correlation experiments to infer the characteristics of multipartite quantum states without any prior knowledge of the entangled state. To illustrate this concept, we use the bipartite Bell-CHSH test as an example. In the standard Bell-CHSH scenario, two observers, Alice and Bob, perform correlation measurements from spacelike-separated laboratories. Each observer has two binary measurement operators, represented as $A_0$ and $A_1$ for Alice, and $B_0$ and $B_1$ for Bob. Physically, the term binary operators signifies that the quantum particles involved possess two energy levels. The Pauli matrices, which represent these binary operators, act as operators in this system and can be interpreted as rotations on the Bloch sphere. Mathematically, binary operators signify that the Pauli matrices or their combinations have eigenvalues restricted to $1$ and $-1$. This reflects the fact that any single measurement produces one of two possible outcomes: either $1$ or $-1$. Moreover, the binary operators are involutory, meaning they square to the identity, satisfying $A_j^2=\mathbb{I}$ and $B_k^2=\mathbb{I}$. This is due to the fact that when a Pauli matrix acts on its eigenvector, it produces the eigenvector multiplied by its corresponding eigenvalue. Since the eigenvalues are either $1$ or $-1$, squaring the matrix restores the eigenvalue to $1$. Last but not least, since the binary operators used by Alice and Bob act on separate subsystems, they commute, which means they can simultaneously have eigenvalues. As such, the bipartite Bell-CHSH inequality is expressed as
\begin{align}\label{Bell}
I_2\equiv \langle A_0B_0\rangle+ 
\langle A_0B_1\rangle+\langle A_1B_0\rangle-\langle A_1B_1 \rangle \leq 2,
\end{align}
where the bracket $\langle \cdots \rangle$ denotes the expectation value of the binary operator products evaluated with respect to the entangled bipartite state shared by the two observers, Alice and Bob. The difference between the standard Bell test and the CHSH test lies in the type of entangled states they utilize. While the standard Bell test specifically employs maximally entangled states, the CHSH test is more versatile, allowing for the use of general arbitrary entangled bipartite states. The maximally violation of Eq.\:\eqref{Bell}, known as Tsirelson's bound, is attained when the entangled state shared by Alice and Bob is the standard Bell state, subject to local unitary transformations. Using the involutory property of each operator and the commutativity of the binary operators employed by Alice and Bob, the maximal violation can be derived succinctly via the sum-of-squares technique
\begin{align*}
    2\sqrt{2}-I_2=\langle(\frac{A_0+A_1}{\sqrt{2}}-B_0)^2\rangle+\langle(\frac{A_0-A_1}{\sqrt{2}}-B_1)^2\rangle
\end{align*}
where the right-hand side is non-negative, signifying that the maximum quantum violation of the Bell-CHSH inequality amounts to $2\sqrt{2}$. Furthermore, achieving this maximal quantum violation requires Alice and Bob's binary operators to conform to the relations $B_0=(A_0+A_1)/\sqrt{2}$ and $B_1=(A_0-A_1)/\sqrt{2}$. This will enforce the anti-commutativity of the binary operators for each party, $A_0A_1+A_1A_0=0$, and likewise for $B_k$. If we define $A_2\equiv -i[A_0,A_1]/2$, then these three operators have the same algebra as the Pauli matrices. Using this algebraic property, one can readily show that these constraints uniquely determine the entangled state to be the maximally entangled Bell state—up to local unitary transformations. Moreover, any slight deviation from the Tsirelson bound results in only minor variations from the Bell state. This robustness is referred to in the literature as quantum rigidity \cite{summers1987maximal}. The principle of quantum self-testing, combined with the property of quantum rigidity, forms the core of our proposal for a quantum-secured Global Navigation Satellite System (GNSS) \cite{gunthner2017quantum}.

We now extend the quantum self-testing framework to encompass the five-partite scenario—an essential step toward achieving a quantum-secured GNSS. In this context, five observers—specifically, four satellites and one receiver—perform correlation measurements on their respective local binary operators from spacelike-separated laboratories. Each local observer conducts two measurements by employing a pair of binary Hermitian operators, $A_0^j$ and $A_1^j$, where $j = \{1, 2, 3, 4, 5\}$ specifies the five observers. Similar to the standard Bell-CHSH scenario, the local binary operators are involutory, satisfying $(A_0^j)^2=(A_1^j)^2=\mathbb{I}$. Moreover, the spacelike-separation condition guarantees that binary operators associated with different observers commute. In what follows, we will establish that the five-qubit code embodies the unique quantum rigidity properties that are intrinsic to quantum-secured GNSS. Here, the five-qubit code is defined as the simultaneous eigenstate of four commuting stabilizer operators, satisfying \( S_k\ket{\psi} = \ket{\psi} \) for \( k \in \{1,2,3,4\} \), where  
\begin{gather}
    S_1 \equiv X_1Z_2Z_3X_4, S_2 \equiv X_2Z_3Z_4X_5,  \nonumber\\
    S_3 \equiv X_3Z_4Z_5X_1,  S_4 \equiv X_4Z_5Z_1X_2,
\end{gather}
Thus, the entangled subspace of the five-qubit error correction code is two-dimensional, spanned by two highly entangled basis states that define the logical qubits \( |0_L\rangle \) and \( |1_L\rangle \) as \cite{gottesman1997stabilizer}
\begin{subequations}\label{eq:lo}
\begin{align}
|0_L\rangle &= \frac{1}{4} ( |00000\rangle + |10010\rangle + |01001\rangle + |10100\rangle \nonumber\\
&+ |01010\rangle 
- |11011\rangle - |00110\rangle - |11000\rangle \nonumber\\
&- |11101\rangle - |00011\rangle 
- |11110\rangle - |01111\rangle \nonumber\\
&- |10001\rangle - |01100\rangle - |10111\rangle + |00101\rangle ),\\
|1_L\rangle &= \frac{1}{4} ( |11111\rangle + |01101\rangle + |10110\rangle + |01011\rangle \nonumber\\
&+ |10101\rangle 
- |00100\rangle - |11001\rangle - |00111\rangle \nonumber\\
&- |00010\rangle - |11100\rangle 
- |00001\rangle - |10000\rangle \nonumber\\
&- |01110\rangle - |10011\rangle - |01000\rangle + |11010\rangle ).
\end{align}
\end{subequations}
Similar to the Bell-CHSH scenario, the observers perform correlation experiments in spacelike-separated laboratories and test the following quantity for the entangled subspace
\cite{baccari2020device}
\begin{align}\label{eq:I5}
    I_5 &=  \braket{(A_0^{1} + A_1^{1}) A_1^{2} A_1^{3} A_0^{4}} +  \braket{A_0^{2} A_1^{3} A_1^{4} A_0^{5}} \nonumber \\
    & +  \braket{(A_0^{1} + A_1^{1}) A_0^{3} A_1^{4} A_1^{5}} + 2\braket{(A_0^{1} - A_1^{1}) A_0^{2} A_0^{4} A_1^{5}}.
\end{align}
The maximal classical value of this quantity is $5$, which can be achieved by choosing a deterministic strategy where all binary measurement observables $A_0^j$ and $A_1^j$ have pre-existing values of $\pm1$. One can also express $I_5$ via a set of pseudo-stabilizers as: $I_5=\sqrt{2}\braket{\tilde{S}_1+\tilde{S}_3}+\braket{\tilde{S}_2}+2\sqrt{2}\braket{\tilde{S}_4}$, where
\begin{align}
     \tilde{S}_1 \equiv \frac{(A_0^{1} + A_1^{1})}{\sqrt{2}} A_1^{2} A_1^{3} A_0^{4}&,  \tilde{S}_2 \equiv A_0^{2} A_1^{3} A_1^{4} A_0^{5},  \nonumber\\
    \tilde{S}_3 \equiv \frac{(A_0^{1} + A_1^{1})}{\sqrt{2}} A_0^{3} A_1^{4} A_1^{5}&,  \tilde{S}_4 \equiv \frac{(A_0^{1} - A_1^{1})}{\sqrt{2}} A_0^{2} A_0^{4} A_1^{5}.
\end{align}
Using the sum-of-squares method \cite{fang2021sum}, one arrives at
\begin{align}\label{FiveQubitBell}
&4\sqrt{2}+1-I_5= \frac{1}{\sqrt{2}}\langle(\mathbb{I}-\tilde{S}_1)^2\rangle\nonumber\\
&+\frac{1}{2}\langle(\mathbb{I}-\tilde{S}_2)^2\rangle+\frac{1}{\sqrt{2}}\langle(\mathbb{I}-\tilde{S}_3)^2\rangle+\sqrt{2}\langle(\mathbb{I}-\tilde{S}_4)^2\rangle.
\end{align}
Since the right-hand side of Eq.\:\eqref{FiveQubitBell} is non-negative, the maximal quantum violation is thus \(4\sqrt{2}+1\). Conversely, if a five-partite state \( |\psi\rangle \) reaches this maximal violation, then it must be the common eigenstate of the pseudo-stabilizers \( \tilde{S}_j \). As such, the five-partite entangled state must be spanned by the logical qubits \( |0_L\rangle \) and \( |1_L\rangle \), up to local unitary transformations and auxiliary states \cite{baccari2020device}. This property is precisely what defines quantum rigidity or self-testing, forming the core of our quantum-secured global positioning \cite{summers1987maximal, mckague2012robust}.

\section{Synchronizing clocks}
We consider a realistic scenario of global positioning involving four satellites, denoted as \( \mathcal{S}_i \) for \( i = 1,2,3,4 \), and a single receiver \( R \), which could be a car, a ship, or an aircraft. Each satellite generates and broadcasts a five-qubit code to the receiver, where one qubit carries information about its spatial coordinates. Upon receiving the code, each satellite records the arrival time of photons according to its atomic clock, while receiver \( R \) logs the reception time using its own clock.

To ensure the integrity of the signal and prevent hacking or jamming, a five-partite Bell measurement is performed. Subsequently, \( R \) computes its distance to each satellite by determining the signal’s travel time, which is obtained by calculating the difference between transmission and reception times, considering the speed of light. The five-qubit code can originate from various sources, including ground-based stations, satellites, or the receiver itself. An efficient design would involve each satellite having its own independent atomic clock, code generator, and transmitter, streamlining the system for enhanced synchronization and reliability in quantum-secured positioning.

Each transmission is accompanied by a series of Bell experiments to certify the code, ensuring its security through the maximal violation of Bell’s inequality, i.e., $I_5=4\sqrt{2}+1$. A total of four rounds of Bell experiments are required to complete a single quantum-secured positioning task—three for spatial coordinates and one for temporal data.

The critical aspect of this process is maintaining high code fidelity to significantly suppress environmental errors. Only when these errors are sufficiently minimized can hacking or jamming be effectively distinguished, and identified. More importantly, an attacker attempting to compromise the system would need to interfere with at least three satellites, an action guaranteed by the error-correction code to expose their position, making detection unavoidable. This strategic advantage forms the core benefit of our protocol, enabling hostile activities to be automatically tracked, and corrected. 

To optimize signal transmission rate, it is crucial to minimize total gate time, transfer time, and transducing time. GPS relies on continuous signal updates to accurately determine position, requiring receivers to lock onto signals from multiple satellites simultaneously and measure their arrival times precisely. If signal updates become less frequent, the receiver may struggle to maintain a reliable fix, particularly in dynamic environments such as a moving vehicle. This can lead to increased latency, causing delays in positioning updates, loss of lock, where receivers lose connection with satellites, and error amplification, which reduces synchronization accuracy and may disrupt system functionality. Ensuring rapid signal transmission and minimizing delays allows GPS systems to maintain accuracy and efficiency even in fast-changing conditions.

Note that our protocol is not limited to the global positioning system. A realistic alternative to it is to deploy a quantum-secured positioning system using low Earth orbit (LEO) satellites \cite{ntanos2021leo, picciariello2024quantum}. For a small-to-medium city (50–100 km wide), a minimum of 4–6 satellites would likely suffice to ensure complete coverage without blind spots and provide reliable service. This approach significantly reduces the stringent requirements for quantum hardware while also mitigating the negative effects of slow code production, which can cause delays in positioning updates. The receiver sends a request requiring a five-qubit code, and the satellites employ an artificial intelligence-assigned computational method to determine which four satellites will share the five-qubit code and synchronize the receiver’s clock.

In the next section, we will examine the detailed experimental plan for the generation, transfer, and transduction of the five-qubit code utilized in our quantum-secured positioning protocol.

\section{Experimental Realization}

To experimentally implement the quantum-secured global positioning scheme, satellites must be capable of generating high-fidelity on-chip multipartite entangled states. We begin this process with a superconducting quantum computing circuit, typically constructed from transmon qubits via Josephson junction designs, to generate a five-qubit quantum error-correcting code. This approach leverages a streamlined quantum circuit architecture, relying exclusively on single-qubit Hadamard gates and two-qubit CNOT gates, which will be further discussed in this section. While the five-qubit code can also be realized using alternative platforms such as trapped-ion qubits, these systems generally exhibit slower operational speeds and lower fidelity compared to superconducting qubits. Later in this section, we provide a detailed evaluation of the total time required to generate the five-qubit code across various quantum computing platforms, based on current experimental data. Within our protocol, this phase is referred to as the generation stage.

Following the generation stage, the next step is on-chip atom–photon quantum-state transfer. In this stage, the quantum state of confined qubits—whether transmon qubits or trapped-atom qubits—is mapped onto on-chip microwave modes, thereby establishing an interface between atomic qubits and flying qubits. For example, each qubit may be coupled to its own microwave resonator, and by applying precisely engineered microwave pulses (using techniques such as Raman transitions or STIRAP), the state of each qubit can be coherently transferred to the corresponding on-chip microwave mode. To date, no widely accepted experimental demonstration has achieved the complete transfer of a multipartite entangled state—such as a GHZ state, a W state, or the five-qubit code discussed here—from atomic qubits to photonic qubits, regardless of the encoding method (e.g., polarization, dual-rail, or time-bin). Most existing research has focused on bipartite Bell-state transfers. However, several theoretical proposals and preliminary studies indicate that multipartite entanglement transfer is technologically feasible. For instance, recent protocols based on Rydberg atoms have suggested viable approaches for transferring W states to photonic states \cite{ramaswamy2025entanglement}. Since the underlying physics of this process is similar to that of bipartite Bell-state transfer, it is reasonable to anticipate that transferring the five-qubit code is also technologically viable. Within our protocol, this phase is referred to as the transfer stage.

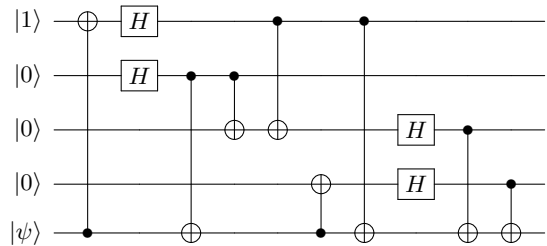
\begin{figure}[tb]
    \centering
    \begin{minipage}{\textwidth}
        \Qcircuit @C=1em @R=1em {
            \lstick{|1\rangle}  & \targ     & \gate{H} & \qw      & \qw      & \ctrl{2} & \qw      & \ctrl{4}  & \qw      & \qw      & \qw      & \qw \\
            \lstick{|0\rangle}  & \qw      & \gate{H} & \ctrl{3} & \ctrl{1} & \qw      & \qw      & \qw      & \qw      & \qw      & \qw      & \qw \\
            \lstick{|0\rangle}  & \qw      & \qw      & \qw      & \targ     & \targ    & \qw      & \qw      & \gate{H}  & \ctrl{2}      & \qw      & \qw \\
            \lstick{|0\rangle}  & \qw      & \qw      & \qw      & \qw      & \qw      & \targ   & \qw      & \gate{H}      & \qw  & \ctrl{1}     & \qw \\
            \lstick{|\psi\rangle} & \ctrl{-4} & \qw    & \targ    & \qw      & \qw      & \ctrl{-1}      & \targ     & \qw     & \targ   & \targ    & \qw
        }
    \end{minipage}
    \caption{A compact quantum circuit designed to generate a five-qubit code, consisting of four Hadamard gates and eight CNOT gates \cite{mondal2024optimized}. The circuit transforms an arbitrary single-qubit state, $\ket{\psi} = \alpha \ket{0} + \beta \ket{1}$, into the encoded state $\alpha \ket{0_L} + \beta \ket{1_L}$ (see Eq \eqref{eq:lo}).}
    \label{Fig:circuit}
\end{figure}

After the transfer stage, the on-chip entangled photons are transduced into flying entangled photons, which are then broadcast to other satellites and users. Since superconducting qubits and on-chip entangled photons operate in the microwave regime (GHz), while flying photons typically function in the terahertz (THz) or optical frequency range, an efficient quantum transducer is required. In this process, quantum transduction—also known as quantum frequency conversion—enables the faithful conversion of entangled photons between different frequency ranges. The substantial frequency gap between microwave and terahertz or optical waves necessitates transduction via intermediate bosons or nonlinear processes. Conventional transduction techniques, such as optomechanics and electro-optics, typically operate at extremely low temperatures—ranging from milli-Kelvin to several Kelvin—depending on the material and experimental conditions. Recently, researchers have explored quantum transduction via ferromagnetic and antiferromagnetic magnons \cite{sekine2024microwave} because these approaches could potentially function at much higher temperatures, even reaching room temperature. However, despite their promise, these alternative methods still grapple with significant challenges, including low transduction efficiency. In a single-layer structure, the estimated efficiency is approximately \(\eta \approx 10^{-9}\). However, by employing heterostructures consisting of roughly 5000 alternating layers of antiferromagnetic materials and nonmagnetic insulators, the efficiency can be significantly enhanced to around \(\eta \approx 10^{-2}\) \cite{sekine2024microwave}. Despite this improvement, theoretical models indicate that a minimum efficiency of \(\eta = 0.5\) is necessary to ensure reliable quantum state transfer between quantum devices. Within our protocol, this phase is referred to as the transducing stage.

Upon the arrival of flying photons, the satellites and receivers can simply reverse the transducing state and perform measurements on the on-chip entangled microwave. Thus, the successful implementation of this quantum-secured global positioning scheme depends on the fidelity of the single-qubit quantum gate and the two-qubit quantum state, as well as the operation time of the gates, the success rate of entanglement transfer, and the efficiency of bidirectional quantum transduction.

Experimental implementations of the five-qubit code have spanned various quantum computing platforms. In the early development phase of quantum computing, nuclear magnetic resonance (NMR) systems demonstrated this code \cite{knill2001benchmarking, zhang2012experimental, souza2011experimental}. More recently, advancements have led to implementations on superconducting circuits \cite{gong2022experimental}, trapped ions \cite{ryan2022implementing, brown2023advances}, and spin qubits \cite{abobeih2022fault}. In the following, we estimate the time required to generate an atomic five-qubit code during the generation stage across mainstream quantum computing platforms. 

In superconducting qubit architectures, studies demonstrate that fluxonium-based gates equipped with the suppressing counter-rotating errors technique enable single-qubit gate fidelities exceeding 99.997\% with gate times as short as 8.2 nanoseconds \cite{rower2024suppressing}. However, fluxonium-based gates typically operate within the tens of nanoseconds range. For instance, studies have demonstrated a fluxonium-based controlled-Z gate achieving 99.89\% fidelity with a gate time of 85 nanoseconds \cite{ding2023high}. Meanwhile, researchers have recently achieved a controlled-Z gate in superconducting circuits with 99.8\% fidelity in 25 nanoseconds by using a combination of a standard transmon and an inductively shunted transmon with opposite anharmonicities. 

Based on these data, we can estimate the total gate time required to implement the five-qubit code during its generation stage on superconducting qubit circuits. In an ideal quantum circuit, if all gates within a layer start and finish simultaneously, with no overlap or idle time between any neighboring layers, then the total gate time can be estimated by $t_{\text{total}} \approx d_{\text{1q}}  t_{\text{1q}} + d_{\text{2q}}  t_{\text{2q}}$, where $d_{\text{1q}}$ and $d_{\text{2q}}$ represent the depths of single-qubit and two-qubit gates respectively, and $t_{\text{1q}}$ and $t_{\text{2q}}$ denote the fixed gate times for single-qubit and two-qubit gates respectively. In Fig.\:\ref{Fig:circuit}, the compact quantum circuit design consists of four Hadamard gates and eight CNOT gates, while the depths for single-qubit and two-qubit gates are $d_{\text{1q}}=2$ and $d_{\text{2q}}=8$ respectively. Using the estimations $t_{\text{1q}} \approx 8.2$ nanoseconds and $t_{\text{2q}} \approx 25$ nanoseconds, the total gate time required to generate a five-qubit code is calculated as $t_{\text{total}} \approx 216.4$ nanoseconds. Therefore, minimizing the two-qubit gate time is crucial for reducing the total gate time, as two-qubit operations are not only more challenging to parallelize but also significantly slower than single-qubit operations. 

Analyzing the fidelity of the entire circuit is more complex, as noise accumulation is not always independent. In particular, errors do not always behave in a simple additive manner, like adding probabilities of independent events. Real devices often experience more intricate forms of noise, such as coherent errors, leakage, cross-talk, or \(1/f\) noise. The fidelity of the entire circuit is bounded above by the product of the individual gate fidelities, $F_{\text{circuit}} \lesssim F_{\text{1q}}^{n_{\text{1q}}} \cdot F_{\text{2q}}^{n_{\text{2q}}}$, provided that the quantum noises are independent and non-Markovian, where \( F_{\text{1q}} \) and \( F_{\text{2q}} \) represent the fidelities of single-qubit and two-qubit gates respectively, and \( n_{\text{1q}} \) and \( n_{\text{2q}} \) are the numbers of single-qubit and two-qubit gates respectively. Using the estimations \( F_{\text{1q}} \approx 99.997\% \) and \( F_{\text{2q}} \approx 99.8\% \), the fidelity of the entire circuit is calculated to be \( F_{\text{circuit}} \approx 98.4\% \). The overall circuit fidelity diminishes exponentially as the number of gates increases. However, since the infidelity of two-qubit gates is typically orders of magnitude greater than that of single-qubit gates, reducing the number of two-qubit gates is essential for minimizing the total circuit infidelity.

In trapped-ion qubit architectures, gate fidelities are typically high, but gate speeds tend to be slower. This limitation arises because trapped-ion qubits rely on phonon-mediated coupling through shared motional modes, requiring precise timing and calibration for two-qubit gates. While methods like utilizing Rydberg interactions \cite{zhang2020submicrosecond} can boost gate speeds, they often come at the cost of fidelity. For example, a study demonstrated single-qubit gates with an exceptionally low error rate of \(1.5 \times 10^{-6}\) on hyperfine atomic clock qubits using \(^{43}\mathrm{Ca}^{+}\) ions in a cryogenic surface trap, executing gates in just 1.32 microseconds \cite{leu2023fast}. For two-qubit operations, researchers achieved an entangling gate in 700 nanoseconds by leveraging strong dipolar interactions between trapped Rydberg ions, enabling the creation of a Bell state with 78\% fidelity \cite{zhang2020submicrosecond}. To improve fidelity, recent studies show that using clock qubits ($^{1}\text{S}_{0} \leftrightarrow ^{3}\text{P}_{0}$) can enable two-qubit entangling gates with a fidelity of 99.62\%, resulting in a Bell state with 97.6\% fidelity and a four-partite GHZ state with 97\% fidelity \cite{finkelstein2024universal, tsai2025benchmarking}. Another report noted average fidelities of 99.9975\% for single-qubit gates and 99.816\% for two-qubit gates, though specific gate durations were not provided \cite{moses2023race}. A more recent study achieved single-qubit gates with a fidelity of 99.99916\% and two-qubit Bell states with a record-high fidelity of 99.97\%, with gate times around 60 microseconds \cite{loschnauer2024scalable}. With the existing technology, single-qubit gates are anticipated to operate within a few microseconds, whereas two-qubit gates typically require several tens of microseconds to execute.

\begin{table}[tbp]
    \centering
    \begin{tabular}{|c|c|c|}
        \hline
        \textbf{} & \textbf{Superconducting Qubits} & \textbf{Trapped-ion Qubits} \\
        \hhline{|=|=|=|}
        \( t_{1q} \) & 8.2 ns & 1.32 $\mu$s \\
        \hline
        \( t_{2q} \) & 25 ns & 60 $\mu$s \\
        \hline
        \( t_{\text{total}} \) & 216.4 ns & 482.64 $\mu$s \\
        \hline
        \( F_{\text{1q}} \) & 99.997\% & 99.99985\% \\
        \hline
        \( F_{\text{2q}} \) & 99.8\% & 99.97\% \\
        \hline
        \( F_{\text{circuit}} \) & 98.4\% & 99.8\% \\
        \hline
    \end{tabular}
    \caption{Comparison of superconducting and trapped-ion qubit architectures in terms of fidelity of the entire quantum circuit and total gate time required for generating a five-qubit code.}
    \label{tab:qubit_comparison}
\end{table}

Based on the estimations \( t_{1q} \approx 1.32 \) microseconds and \( t_{2q} \approx 60 \) microseconds, the total gate time required for generating a five-qubit code in trapped-ion architectures is calculated as $t_{\text{total}} \approx 482.64 \text{ microseconds}$. In the current setting, one can see from the difference between 482.64 microseconds and 216.4 nanoseconds that, due to the slow two-qubit gate operations in trapped-ion architectures, the total gate time required to generate a five-qubit code is longer compared to superconducting qubit architectures, differing by approximately three orders of magnitude. Besides, using the estimations \( F_{\text{1q}} \approx 99.99985\% \) and \( F_{\text{2q}} \approx 99.97\% \), the fidelity of the entire circuit is calculated to be approximately \( F_{\text{circuit}} \approx 99.8\% \). This implies that the overall infidelity for generating a five-qubit code via the best trapped-ion platform is approximately one-eighth that of the best superconducting platform. 

Based on the above analysis (see Table \ref{tab:qubit_comparison}), we can conclude that when fidelity is the primary concern—particularly in quantum-secured communication—the trapped-ion platform is the preferred choice. This platform, especially when utilizing clock qubits, offers an additional advantage: compatibility with existing atomic clock technology used in GPS systems. Integrating it would require only the addition of a cryogenic cooling system. However, when positioning precision is the main priority, the fastest available platform—such as superconducting systems—should be adopted, provided that fidelity remains above a critical threshold, such as 99\%. This is essential because even minor time delays can cause deviations in distance measurements.

\section{Discussion and Conclusion}
In today’s GPS system, 31 operational navigation satellites orbit Earth at an altitude of approximately 20,200 kilometers, ensuring uninterrupted global signal coverage. However, this immense distance poses a significant challenge for modern quantum technology. To date, the longest-distance Bell experiment involved the Micius satellite, distributing entangled photons between ground stations over 1,200 kilometers apart, with the satellite orbiting at roughly 500 kilometers altitude \cite{yin2017satellite}. At such distances, photon entanglement is susceptible to decoherence from environmental interactions. Conducting a bipartite photon Bell test across 20,200 kilometers remains a major technological barrier, requiring advances in photon transmission, quantum memory, and error correction techniques. 

The Micius experiment's achievement of 1,200 kilometers indicates that multipartite entangled states, such as the five-qubit code, could theoretically be transmitted over similar distances. By utilizing an entangled source aboard a satellite and distributing it to multiple ground stations, this becomes feasible. The vacuum of space significantly reduces loss and decoherence, enabling transmission across distances ranging from hundreds to thousands of kilometers. However, the main challenge lies in generating and detecting multipartite entangled optical states with sufficient fidelity over these vast distances.

While preparing this manuscript, the authors observed that a research team in China had successfully transmitted a Bell state across a distance of 10,000 km using micro-satellites weighing approximately 1 kg each \cite{li2025microsatellite}. This breakthrough brings the concept closer to practical implementation, demonstrating the viability of long-distance quantum communication. Such advancements could lead to the development of a quantum-secured, device-independent global positioning system, enhancing security and precision in navigation technology.

Our protocol offers the advantage that the five-qubit code, shared among the five parties, cannot be copied, as dictated by the no-cloning theorem, nor can it be forged, due to the quantum rigidity property. Furthermore, if a cyberattack is identified, the position of the attacker can be located and tracked. This makes our protocol a promising candidate for a next-generation, quantum-secured global positioning system, delivering cutting-edge advantages in resilient positioning, navigation, and timing for future applications.

\section{Acknowledgments}
EJK acknowledges the support from National Yang Ming Chiao Tung University.

\end{document}